\pdfoutput=1
\documentclass[aps,prl,reprint,twocolumn]{revtex4-1}
\usepackage{graphicx}
\usepackage{amsmath}
\usepackage{epstopdf}
\usepackage{xcolor}

\newcommand{\mne}{Mn$_\mathrm{E}$}
\newcommand{\mnsb}{Mn$_\mathrm{Sb}$}
\newcommand{\sbmn}{Sb$_\mathrm{Mn}$}
\newcommand{\mnni}{Mn$_\mathrm{Ni}$}

\begin{document}

\title{Spectral signatures of thermal spin disorder and excess Mn in half-metallic NiMnSb}

\author{K.\ D.\ Belashchenko}

\author{Jeevaka Weerasinghe}

\author{Sai Mu}

\author{B.\ S.\ Pujari}

\altaffiliation[Present address: ]{Centre for Modeling and Simulation, Savitribai Phule University of Pune, Ganeshkhind, Pune 411007, India}

\affiliation{Department of Physics and Astronomy and Nebraska Center for Materials and Nanoscience, University of Nebraska-Lincoln, Lincoln, Nebraska 68588, USA}

\date{\today}

\begin{abstract}
Effects of thermal spin disorder and excess Mn on the electronic spectrum of half-metallic NiMnSb are studied using first-principles calculations.
Temperature-dependent spin disorder, introduced within the vector disordered local moment model, causes the valence band at the $\Gamma$ point to broaden and shift upwards, crossing the Fermi level and thereby closing the half-metallic gap above room temperature. The spectroscopic signatures of excess Mn on the Ni, Sb, and empty sites (Mn$_\mathrm{Ni}$, Mn$_\mathrm{Sb}$, and Mn$_\mathrm{E}$) are analyzed. Mn$_\mathrm{Ni}$ is spectroscopically invisible. The relatively weak coupling of Mn$_\mathrm{Sb}$ and Mn$_\mathrm{E}$ spins to the host strongly deviates from the Heisenberg model, and the spin of Mn$_\mathrm{E}$ is canted in the ground state. While the half-metallic gap is preserved in the collinear ground state of Mn$_\mathrm{Sb}$, thermal spin disorder of the weakly coupled Mn$_\mathrm{Sb}$ spins destroys it at low temperatures. This property of \mnsb\ may be the source of the observed low-temperature transport anomalies.
\end{abstract}

\maketitle

Half-metallic ferromagnets \cite{Groot,Galanakis,Felser,Katsnelson} with a band gap in one spin channel can serve as sources of highly spin-polarized currents in spintronic devices \cite{Zutic,BGW}. While many materials, including a number of Heusler compounds \cite{Galanakis,Felser}, are predicted by band structure calculations to be half-metallic, experimental verification is complicated by defects, thermal spin excitations, and surface non-stoichiometry \cite{Dowben}. Correlation effects can also lead to spin depolarization already at zero temperature \cite{Chioncel}.

If a perfect crystal is half-metallic at $T=0$, there may be a region of low $T$ and weak disorder where transport is effectively single-channel \cite{BGW}. However, a crossover to two-channel \cite{Mott} conduction should occur if a quasiparticle band crosses the Fermi level $E_F$ in the gapped spin channel. Thermally-induced incoherent spectral density in the gap can also lead to transport anomalies.

A bulk-sensitive positron annihilation measurement \cite{Hanssen} found NiMnSb to be half-metallic to experimental accuracy, but other measurements paint a more complicated picture. An anomaly in the transport properties was observed near 80 K \cite{Hordequin,Borca}. It was suggested \cite{Lezaic} that it could be due to the longitudinal fluctuations of the Ni spin moments, but the effect was overestimated by improper phase space integration \cite{Lezaic2}. Recent studies indicate that rather than being intrinsic, the 80 K anomaly is only found in samples with excess manganese \cite{Wang,Zhang}. This suggests the anomaly is not related to the non-quasiparticle states \cite{Chioncel}, which also lack a clear parameter of order 10 meV associated with them. It was suggested that thermal excitations on the interstitial \mne\ atoms may be involved \cite{Zhang}. Here we study the influence of thermal spin fluctuations and excess Mn on the single-particle electronic spectrum of NiMnSb and identify \mnsb\ as the most likely source of the 80 K anomaly.

For averaging over spin disorder in electronic structure calculations, we used the adiabatic disordered local moment (DLM) method \cite{Gyorffy}, which provides excellent agreement with supercell averaging for the density of states (DOS) and electric resistivity of ferromagnetic metals \cite{Glasbrenner,Kudr}. The effects of spin disorder at low temperatures can not be described by the binary pseudo-alloy approximation with collinear ``up'' and ``down'' atomic spin moments \cite{Akai}, which is justified only for the paramagnetic state \cite{Gyorffy}. Instead, we employ proper averaging \cite{Staunton} over all directions of the Mn spins. Below we refer to this approach as the vector DLM (VDLM) model.

In ideal NiMnSb we used the single-site mean-field distribution function $p(\theta)\propto\exp(\alpha\cos\theta)$ for Mn spin directions. The parameter $\alpha=H_W \mu/T$, where $H_W$ is the Weiss field and $\mu$ the local spin moment, is mapped to the temperature using the experimental $M(T)$ curve \cite{Otto,Ritchie}. Both spin disorder and excess Mn defects are treated within the coherent potential approximation (CPA) applied within the Green's function-based linear muffin-tin orbital (GF-LMTO) method \cite{Gunnarsson,Turek,Fe8N,basis}. The azimuthal angle $\phi$ in the CPA equations is integrated out analytically, while the $\theta$ dependence is discretized using the Gauss-Legendre quadrature.
The potentials for all atoms (and $\theta$ angles) are determined self-consistently by embedding the CPA self-consistency loop into the density-functional iteration. To enforce magnetic self-consistency, a transverse constraining field \cite{Stocks} is introduced for each $\theta$. The experimental lattice constant of 5.92 \AA\ and the generalized gradient approximation \cite{PBE} are used throughout.

The electronic structure in the presence of disorder is represented by the energy- and $\mathbf{k}$-resolved CPA spectral function \cite{Turek}. It (and DOS) is resolved by spin in the global reference frame and is accessible through spectroscopic measurements like spin-polarized angular-resolved photoemission (ARPES). Our focus will be on the minority-spin spectral function, which is gapped at $T=0$ in the ideal crystal. A small $10^{-4}$ Ry imaginary part is added to energy to improve CPA convergence; its effect is negligible compared with disorder broadening.

\begin{figure*}[htb]
\centering
\includegraphics[width=0.3\textwidth]{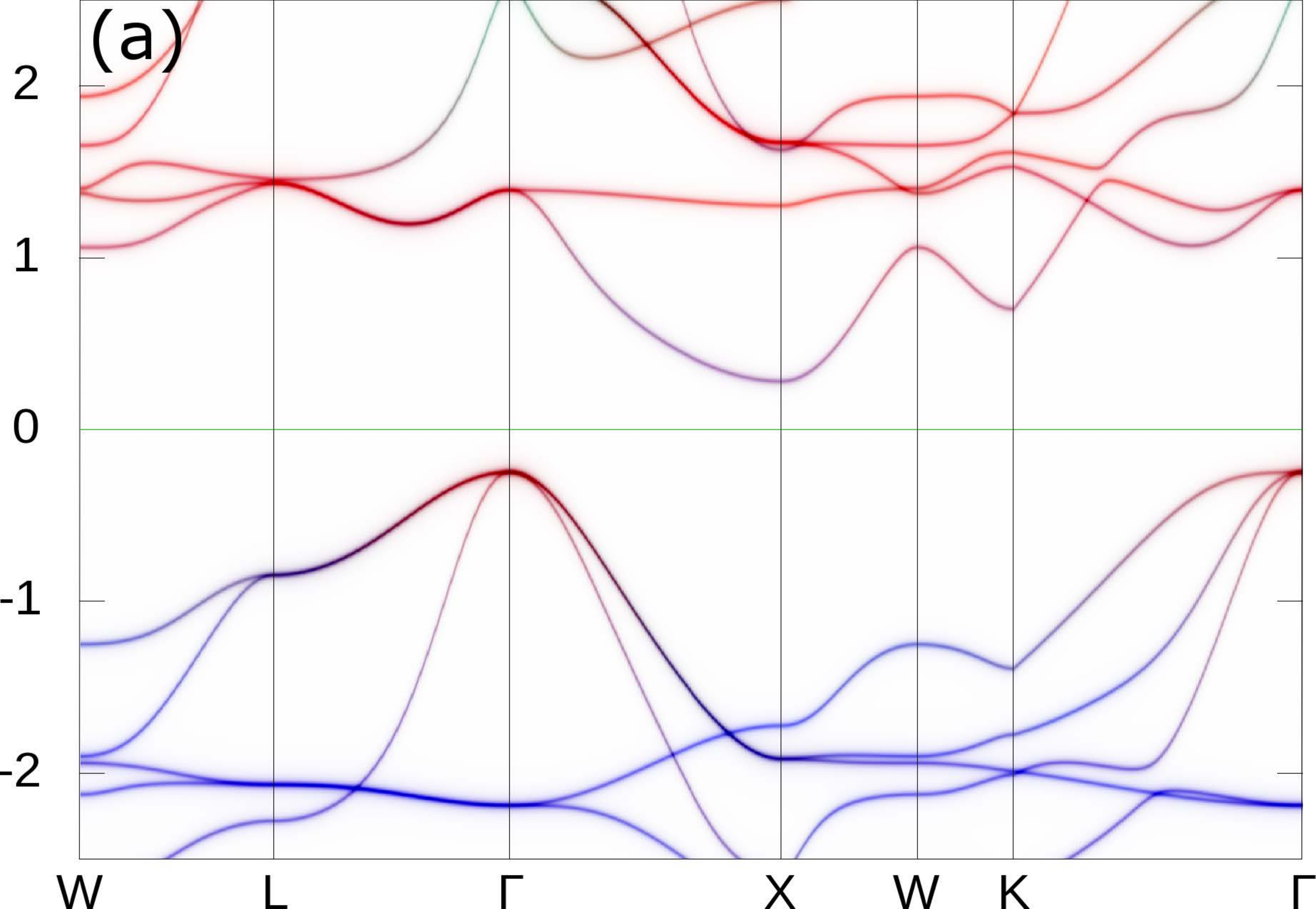}\hfil\includegraphics[width=0.3\textwidth]{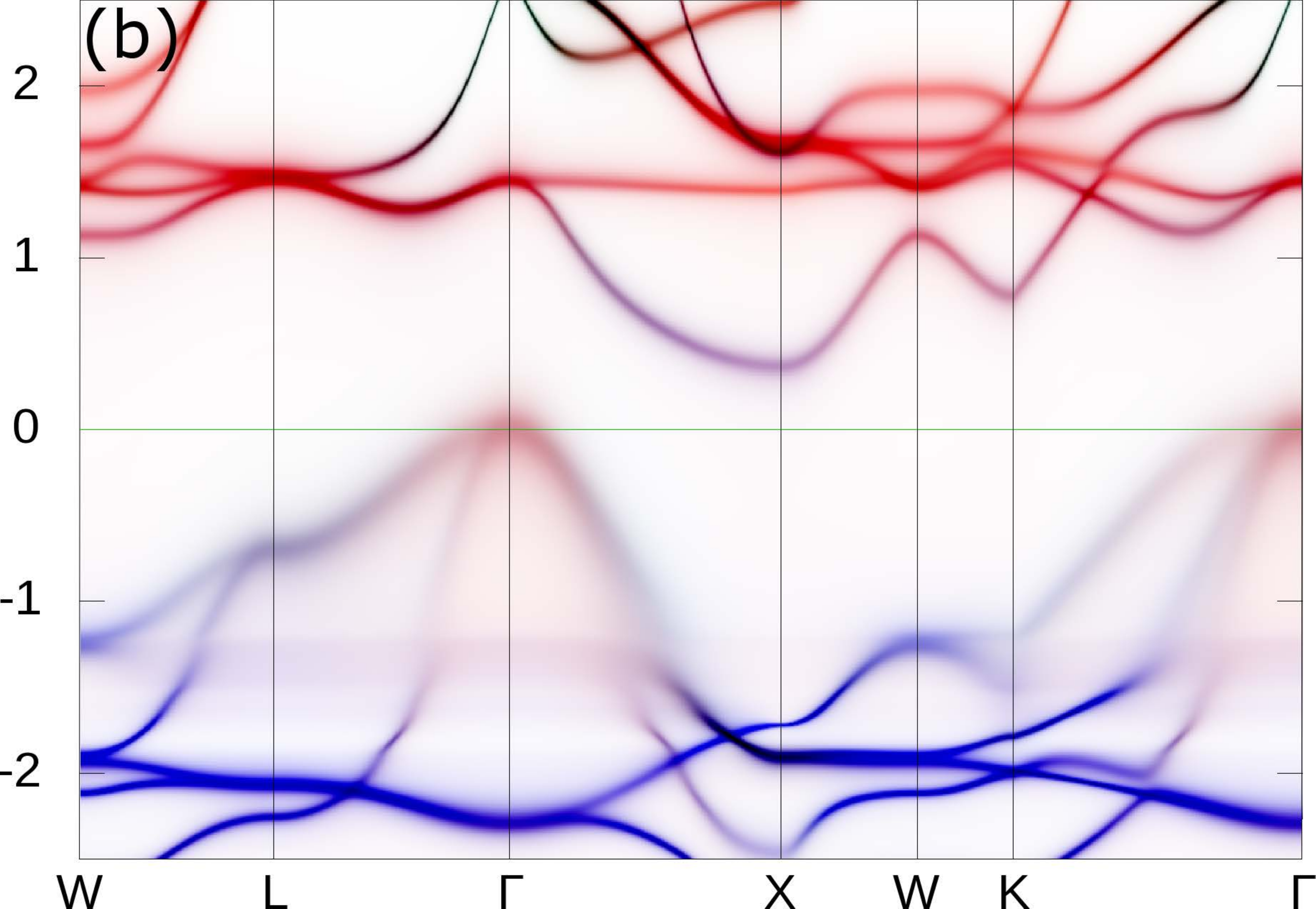}\hfil\includegraphics[width=0.3\textwidth]{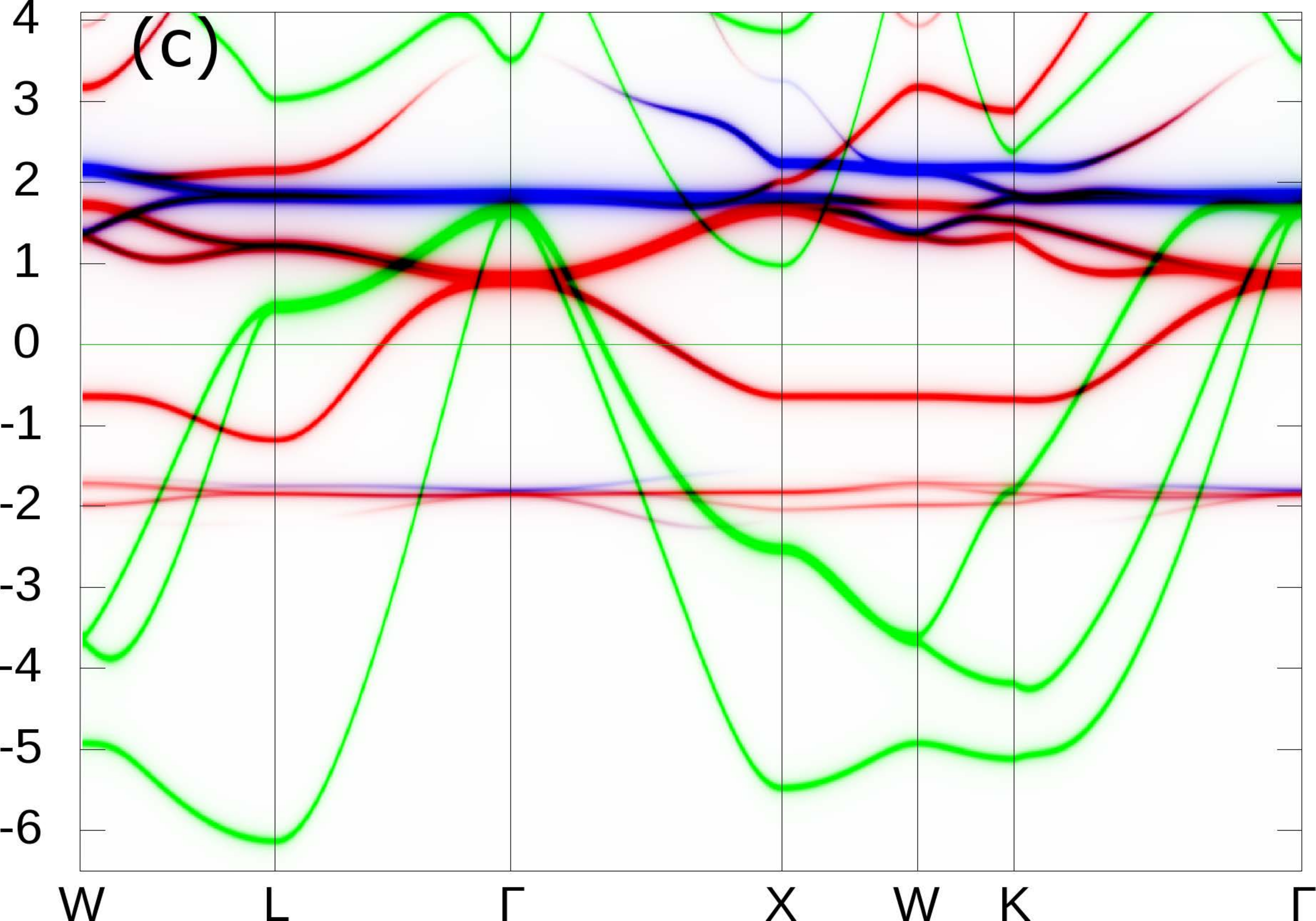}
\caption{(a-b) Minority-spin spectral functions in NiMnSb. (a) $T=0$. (b) Spin disorder (VDLM) at $M(T)/M(0)=0.80$. The coloring reflects the spectral weight contributions of Mn (red), Ni (blue), and Sb (green, barely seen). (c) Hybridization analysis: Red and blue coloring show, respectively, Mn $t_{2g}$ and Mn $e_g$ orbital character when all Sb states except $5s$ are removed from the basis set. Green coloring shows the Sb character when the $3d$ states of Mn and Ni are removed from the basis set. (All charges are taken from the calculation with the full basis set.) Energy in all panels is in eV measured from $E_F$.}
\label{fig:bulk}
\end{figure*}

First we consider the spectroscopic effects of thermal spin disorder in defect-free NiMnSb. Fig.\ \ref{fig:bulk}(a) shows the minority-spin band structure at $T=0$, and Fig.\ \ref{fig:bulk}(b) the spectral function at an elevated temperature corresponding to the reduced magnetization $M(T)/M(0)=0.80$.

Mixing with majority-spin bands is apparent at 1.2-1.6 eV below $E_F$, where those bands and avoided crossings show up in Fig.\ \ref{fig:bulk}(b). The bands that tend to acquire a significant admixture of the opposite spin character at $T\neq0$ are the flat bands with small dispersion. We do not, however, see any trace of the dispersive majority-spin bands that cross the Fermi level. The spectral weight is, of course, transferred from those bands to the minority spin, but this transferred weight is incoherent and spreads over the entire bandwidth. Bloch electrons in dispersive bands behave according to the Stoner picture and ``observe'' an averaged effective field, because adjusting their wavefunctions locally to the directions of individual spins costs a lot of kinetic energy. This general property is favorable for spintronic applications, because the quasiparticles near $E_F$ retain their spin polarization in the presence of spin disorder.

\begin{figure}[htb]
\centering
\includegraphics[width=0.85\columnwidth]{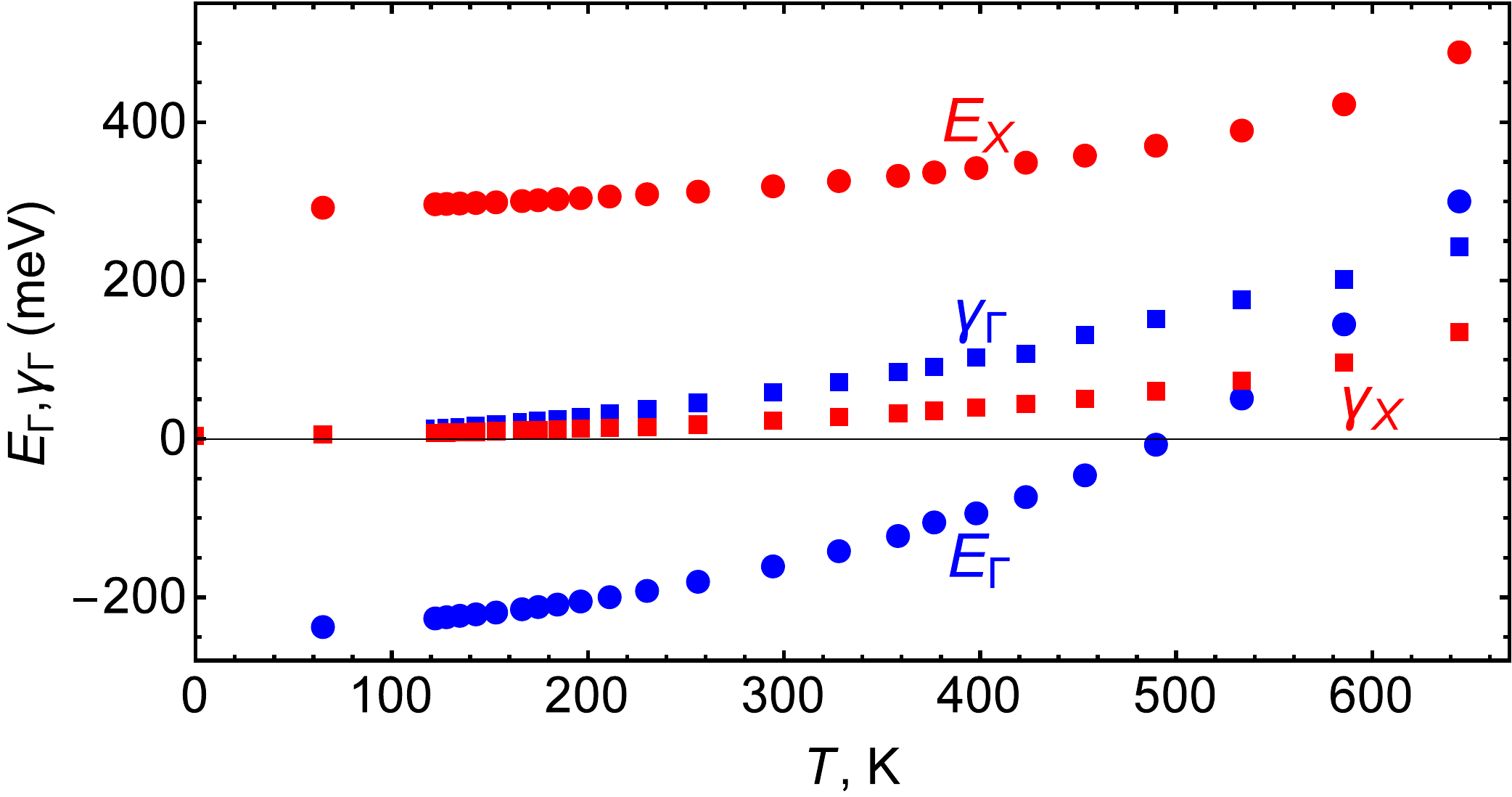}
\caption{Temperature dependence of the band center and half-width of the minority-spin bands at $\Gamma$ and X points at the edges of the half-metallic gap in NiMnSb.}
\label{fig:gandx}
\end{figure}

Fig.\ \ref{fig:bulk}(b) shows that the valence-band maximum (VBM) triplet at $\Gamma$ is strongly broadened and shifted upward by spin disorder. The conduction band minimum (CBM) at X is also broadened and shifted upward but at a smaller rate. Fig.\ \ref{fig:gandx} plots the shift and broadening of VBM and CBM as a function of $T$. The VBM comes within its half-width of $E_F$ at about 400 K. This is where strong transport spin depolarization and crossover to conventional two-channel conduction can be expected. Although this temperature depends on the initial position of $E_F$ at $T=0$, it appears that spin disorder can not lead to low-temperature anomalies in ideal NiMnSb.

The spin disorder-induced \emph{upward} shift of the \emph{minority-spin} VBM states is contrary to the expectation based on the simple exchange splitting picture. This counterintuitive band shift can be understood by examining the hybridization that forms this band in Fig.\ \ref{fig:bulk}(c). As seen in Fig.\ \ref{fig:bulk}(a) (and confirmed by full-potential calculations), the VBM triplet at $\Gamma$ is predominantly composed of Mn $t_{2g}$ states with an admixture of Sb $5p$ and Ni $t_{2g}$ states. If the $3d$ states of Mn and Ni are removed from the basis set, the remaining bands deriving primarily from Sb $5p$ states (green in Fig.\ \ref{fig:bulk}(c)) extend from $-6$ to 2 eV. The $\Gamma$ point is antibonding for these bands. If, on the other hand, all Sb states except $5s$ are removed from the basis set, we end up with the bands for which the Mn character is shown by red ($t_{2g}$) and blue ($e_g$) color in Fig.\ \ref{fig:bulk}(c). Comparing with Fig.\ \ref{fig:bulk}(a), we see that the hybridization of the Mn $t_{2g}$ and Sb $5p$-derived states \cite{Groot} opens the gap and forms the VBM at $\Gamma$. Spin disorder reduces this hybridization by adding the spin overlap factors to the hopping matrix elements, resulting in the upward shift of the VBM triplet seen in Fig.\ \ref{fig:bulk}(b).

As the 80 K anomaly may be due to excess Mn \cite{Wang,Zhang}, we now consider the effects of excess Mn on the Ni, Sb, and empty sites (\mnni, \mnsb, \mne). Various defects in NiMnSb were studied using supercell calculations \cite{Alling,Alling2} and CPA \cite{Orgassa,Alling2}. While the energies of atom-pair swaps \cite{Alling} are related to the formation energies of different defects, the concentration of each non-stoichiometric defect depends on the chemical potentials of the relevant elements and thus on the material synthesis protocol. A polarized neutron diffraction measurement \cite{Brown} of a low-quality crystal found a large concentration of \mnsb\ and \sbmn\ defects. Since the Mn-Sb swap energy is rather high \cite{Alling}, the formation of these defects could be due to nonequilibrium growth conditions.

\begin{figure}[htb]
\includegraphics[width=0.85\columnwidth]{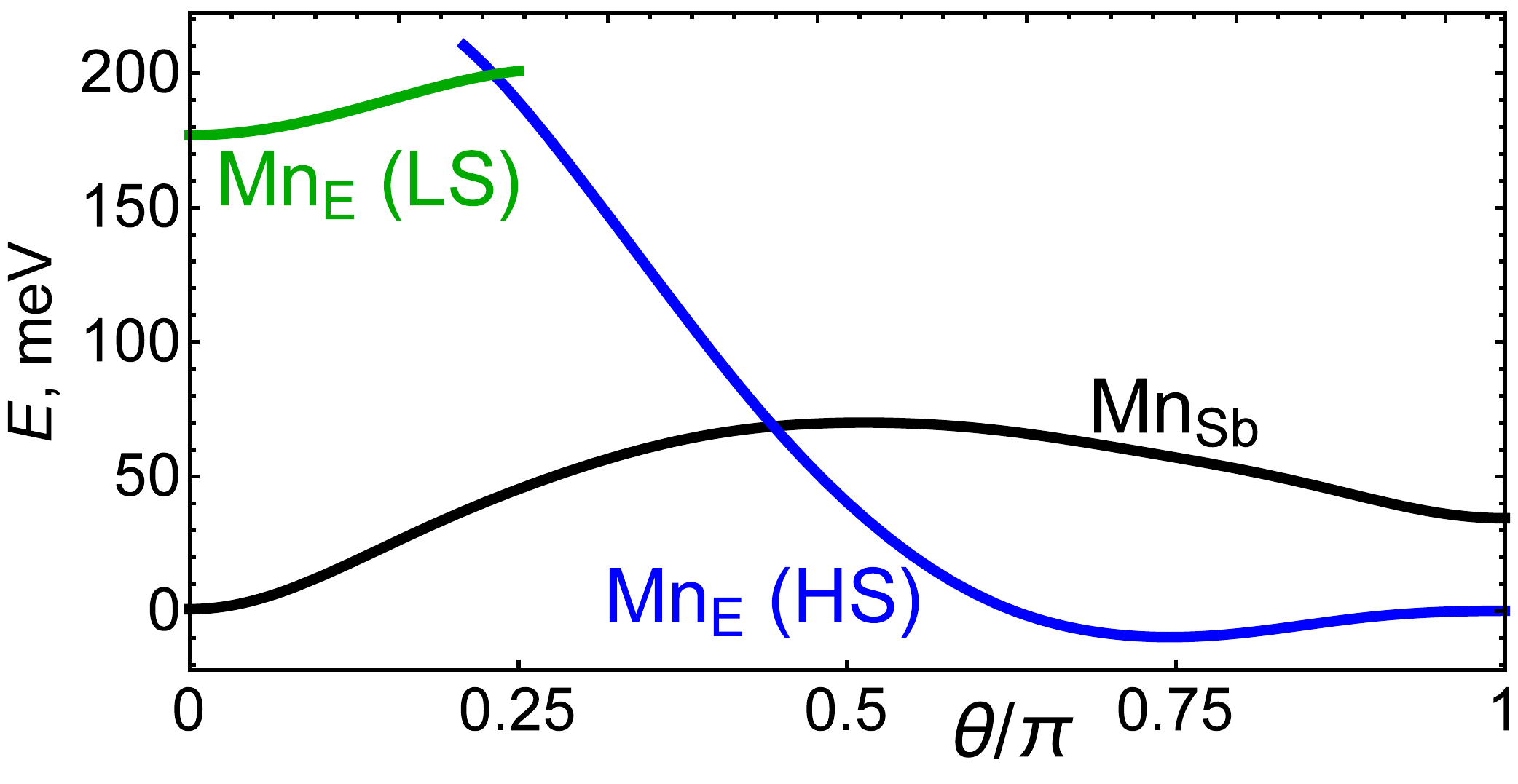}
\caption{Energy of a supercell with one \mne\ or \mnsb\ defect as a function of its spin orientation $\theta$ relative to the host magnetization (from VASP). HS, LS: high spin, low spin.}
\label{fig:rotation}
\end{figure}

We assume excess Mn is randomly distributed over the given sublattice and treat each type of defect separately within CPA. First we used VASP \cite{VASP} to relax the atomic structure for a 48-atom ($2\times2\times1$) host supercell \cite{note} with one excess Mn defect and examine its magnetic coupling to the bulk. For \mnni\ the antiparallel spin alignment is very stable in agreement with Ref.\ \onlinecite{Ekholm}, while the self-consistent solution for the parallel alignment could not be obtained. Fig.\ \ref{fig:rotation} shows the total energy $E(\theta)$ as a function of the angle $\theta$ made by the spin of \mne\ or \mnsb\ with the magnetization. In both cases there are strong deviations from the Heisenberg model predicting $E(\theta)\propto\cos\theta$. For \mne\ we confirm the earlier result \cite{Zhang} that $E(\pi)$ is lower than $E(0)$. However, the energy minimum is reached for a canted spin at $\theta\approx 130^\circ$. There is a high-to-low spin crossover near $\theta=\pi/4$. For \mnsb\ the minimum is at $\theta=0$, but the range of energy variation is only 70 meV.

Since the magnetic coupling of \mne\ and \mnsb\ spins to the bulk is weak, their disordering can lead to anomalies at relatively low temperatures. We now use CPA to study the spectral properties of these defects. We start with the 6.25\% defect concentration as in the VASP supercells, which is also typical for low-quality samples \cite{Brown}.

\begin{figure}[htb]
\includegraphics[width=0.85\columnwidth]{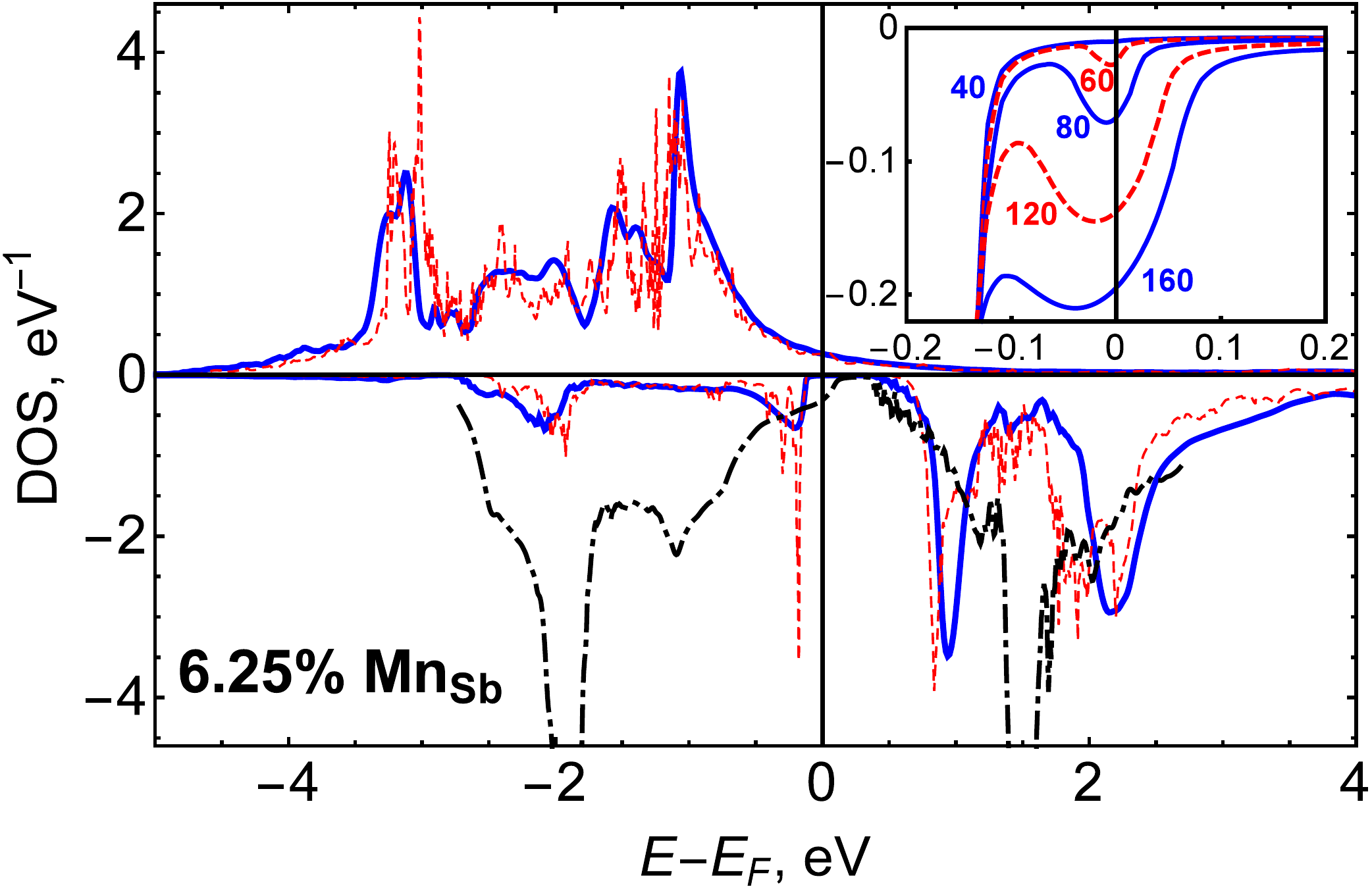}
\caption{Partial DOS for \mnsb\ in NiMnSb with a 6.25\% occupation of the Sb site by Mn, parallel spin alignment. Solid (blue) lines: CPA; dashed (red) lines: VASP. Black dot-dashed line: total minority-spin DOS per formula unit ($N_\downarrow$), disordered spin of \mnsb. Inset: $N_\downarrow$ at 40 K, 60 K, 80 K, 120 K, and 160 K as labeled (same units on axes).}
\label{fig:mn-sb-dos}
\end{figure}

First we performed an LMTO calculation for each supercell, taking the optimized atomic positions from VASP and setting the sphere radius of excess Mn to the same value as the substituted atom \cite{basis}. The partial DOS (PDOS) in these calculations are similar to VASP results, and they give us the correct charge and spin moments in the atomic sphere for the excess Mn atom. In CPA calculations with the ideal lattice, a screening correction to the Madelung potential \cite{scrCPA} is applied with a coefficient that is adjusted to provide the correct charge on Mn. In addition, the local part of the effective GGA magnetic field for the \mne\ defect was scaled by a factor 1.2 to obtain good agreement with the correct local spin moment for \mne\ at both $\theta=0$ and $\pi$. This correction compensates for the lack of atomic relaxations in CPA. The PDOS with these corrections agree well with supercell calculations up to disorder broadening (see Fig.\ \ref{fig:mn-sb-dos} for \mnsb\ at $\theta=0$; the agreement is also good for $\theta=\pi$).

The spectral function with \mnni\ defects (not shown) is only slightly modified compared to ideal NiMnSb (Fig.\ \ref{fig:bulk}(a)), and the broadening of the bands is very small. Since the spin of \mnni\ is strongly coupled to the host magnetization \cite{Ekholm}, this defect is spectroscopically invisible and can not lead to low-temperature anomalies.

\begin{figure*}[htb]
\includegraphics[width=0.82\columnwidth]{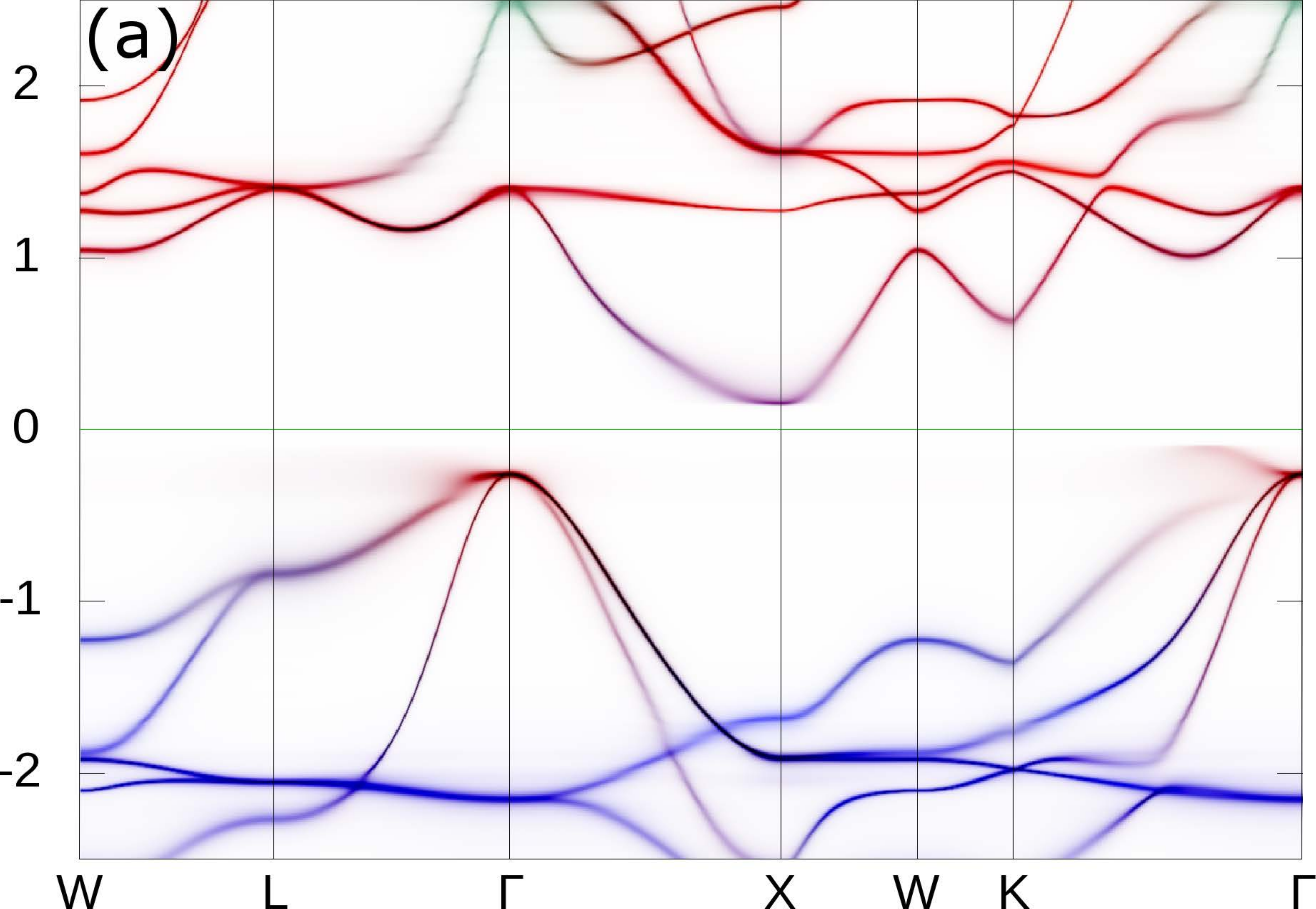}\hfil\includegraphics[width=0.82\columnwidth]{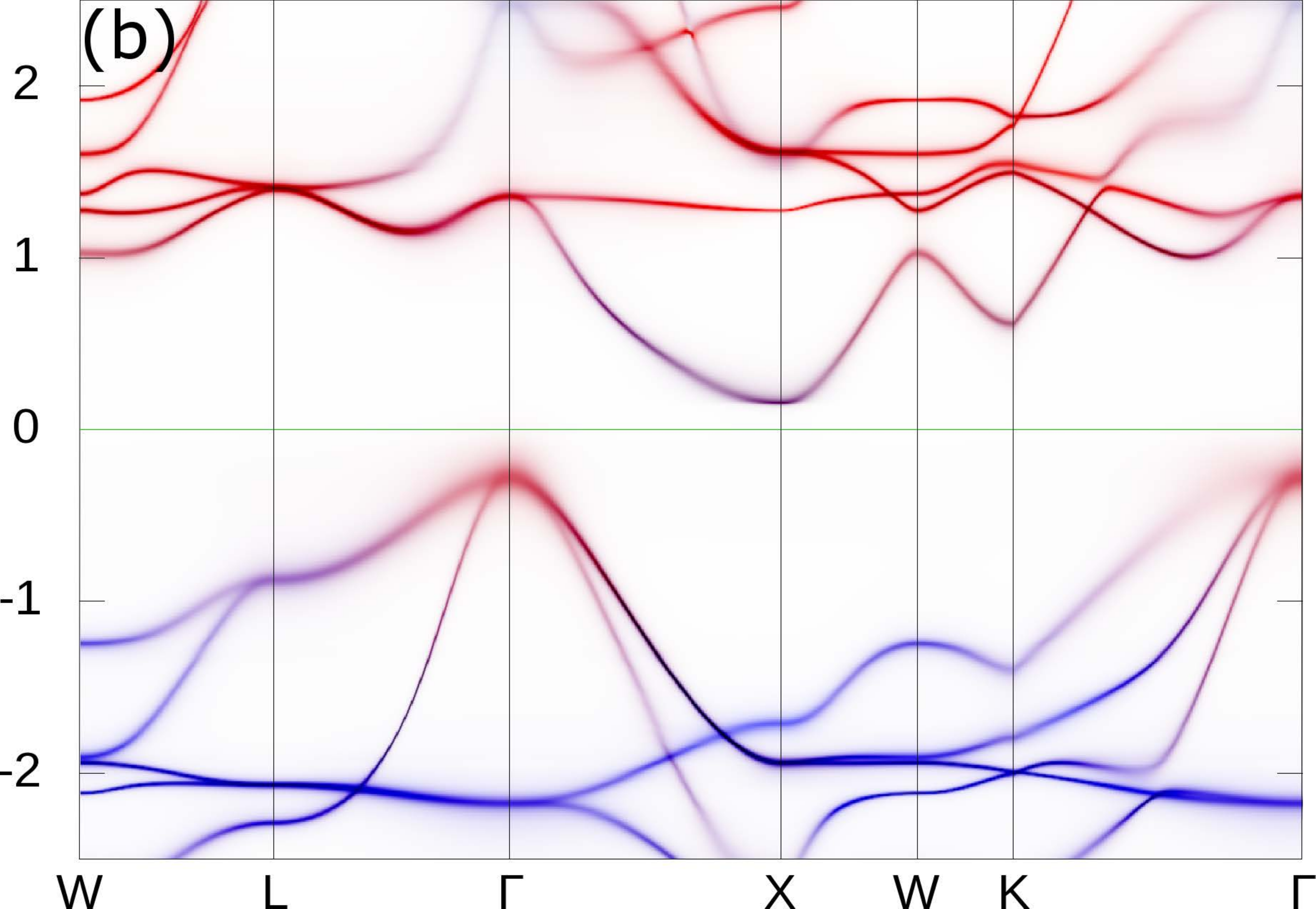}\\
\vskip2ex
\includegraphics[width=0.82\columnwidth]{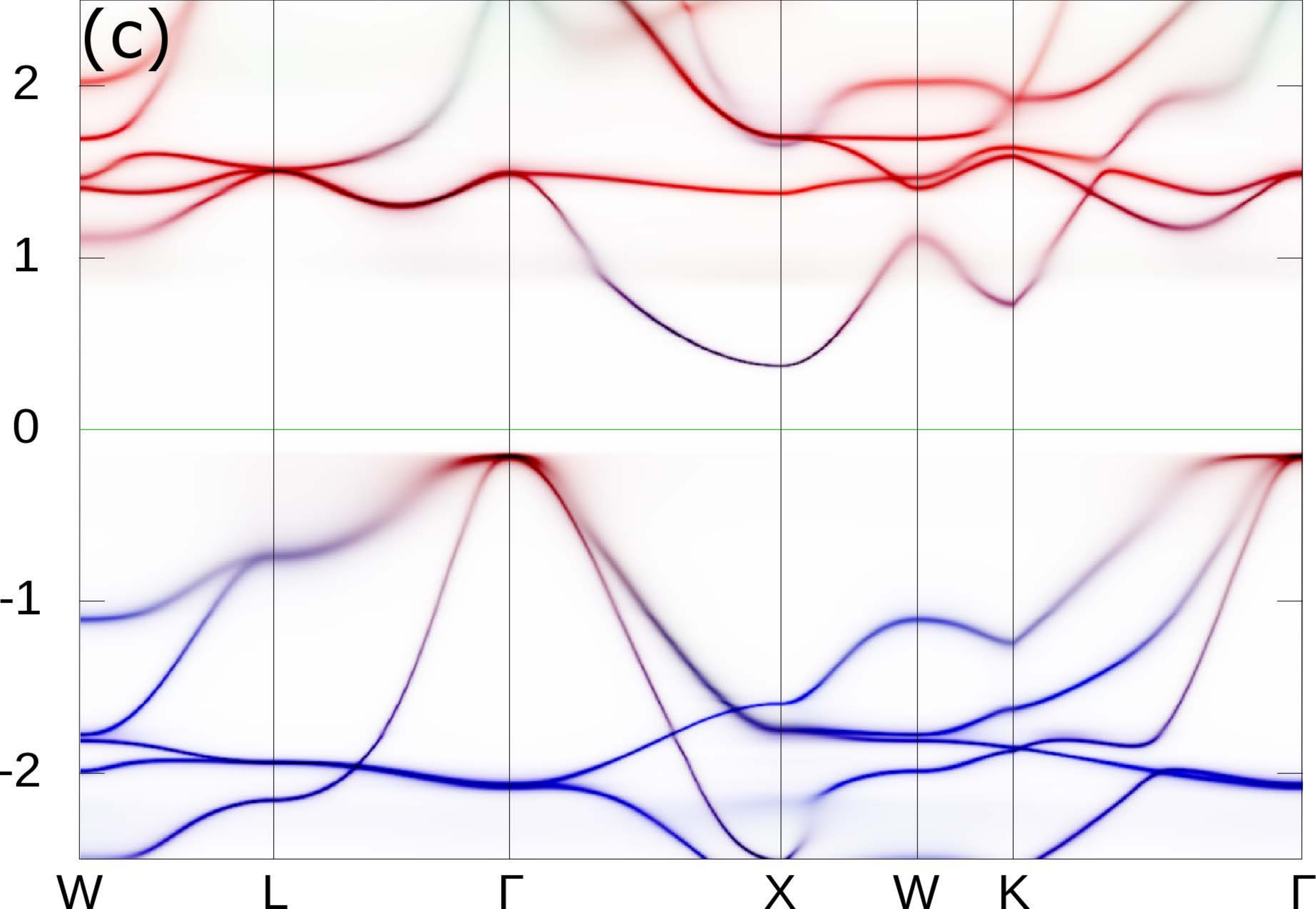}\hfil\includegraphics[width=0.82\columnwidth]{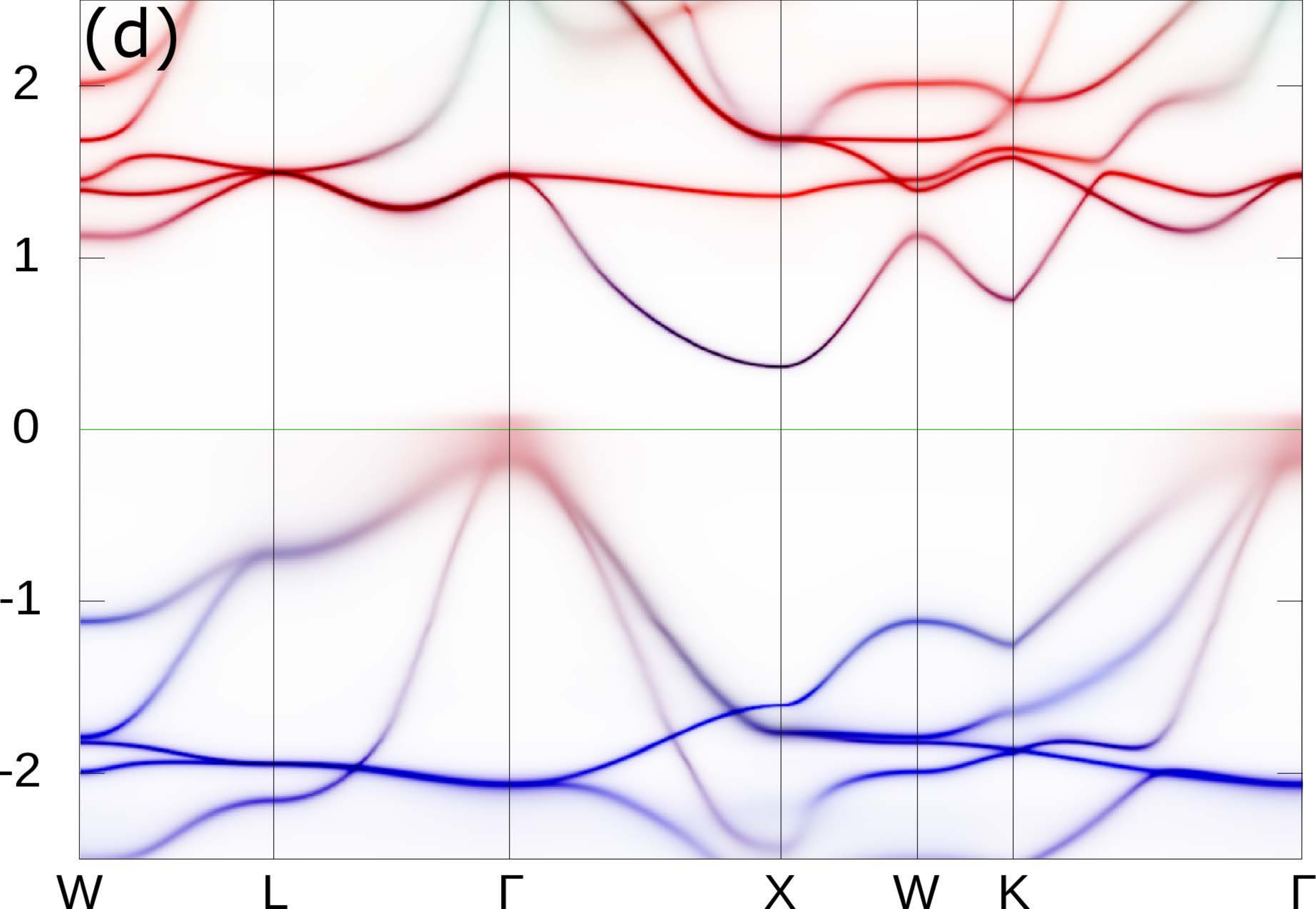}
\caption{Minority-spin spectral function in NiMnSb with 6.25\% of: (a-b) \mne, (c-d) \mnsb.
The spins of \mne\ are: (a) antiparallel to magnetization; (b) randomly oriented.
The spins of \mnsb\ are: (c) parallel to magnetization; (d) randomly oriented.
Coloring is similar to Fig.\ \ref{fig:bulk}(a); \mne\ added to Mn (red), \mnsb\ to Sb (green). Energy is in eV measured from $E_F$.}
\label{fig:mn-es}
\end{figure*}

\begin{figure}[htb]
\includegraphics[width=0.85\columnwidth]{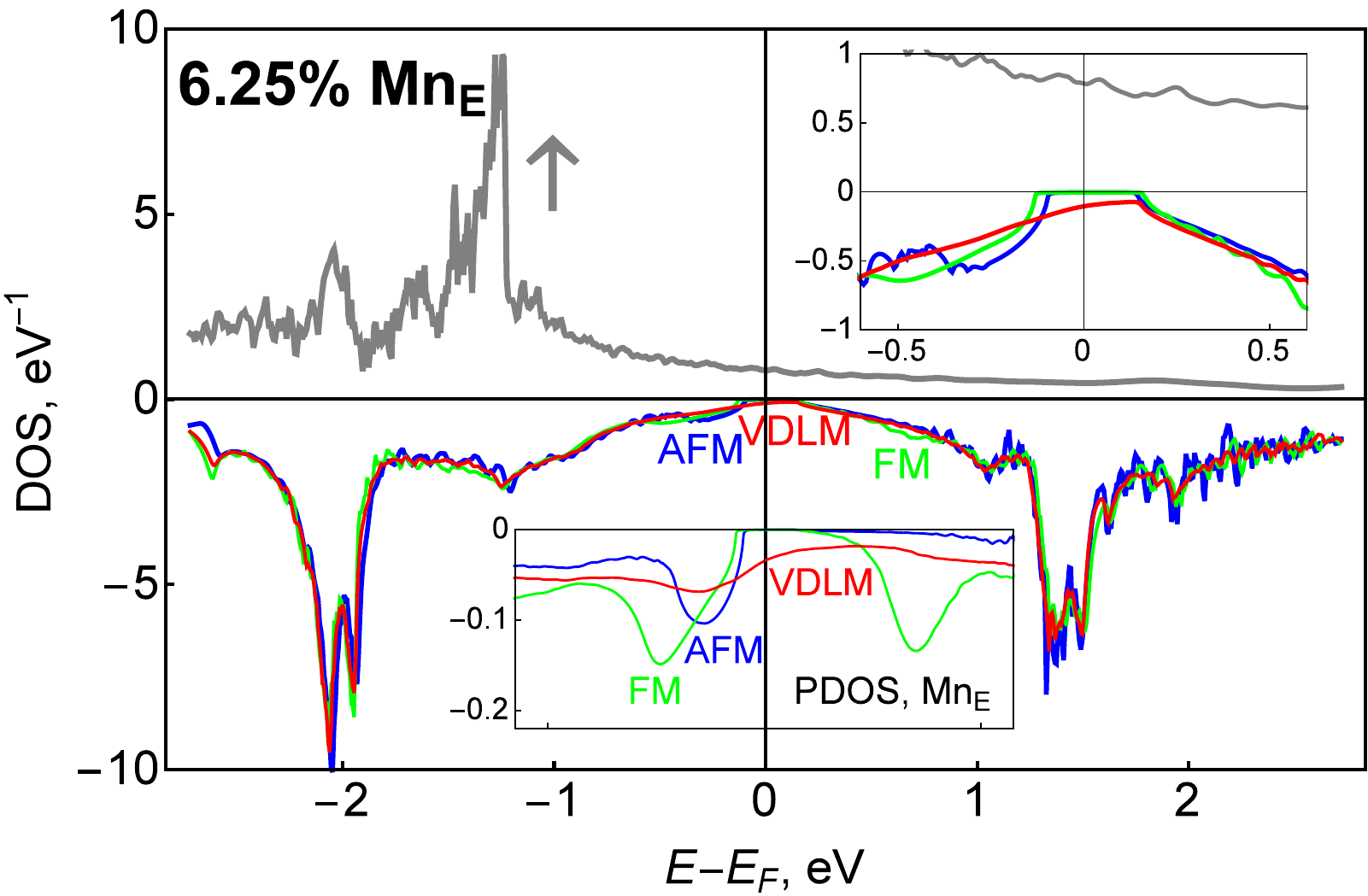}
\caption{Total DOS for NiMnSb with a 6.25\% occupation of the empty site by Mn (CPA). FM (AFM): \mne\ is parallel (antiparallel) to the magnetization. VDLM: fully disordered spin on \mne. Upper inset: magnified region near $E_F$. Lower inset: partial DOS for \mne. (Same units on axes in insets.)}
\label{fig:mn-es-dos}
\end{figure}

Spectral functions for \mne\ at $\theta=\pi$ and in the fully spin-disordered (SD) state are shown in Fig.\ \ref{fig:mn-es}(a-b). For the SD state the $\theta$ integration in CPA is combined with full charge self-consistency. The spin moment of \mne\ varies from 0.84 to 3.04 $\mu_B$ between the points closest to 0 and $\pi$ in good agreement with VASP calculations. PDOS at $\theta=0$, $\pi$, and SD state are shown in Fig.\ \ref{fig:mn-es-dos}.

Fig.\ \ref{fig:mn-es}(a-b) and \ref{fig:mn-es-dos} show that spin disorder on \mne\ introduces a substantial minority-spin spectral weight $N_\downarrow$ inside the half-metallic gap. Its magnitude at $E_F$ is about 13\% of the majority-spin DOS $N_\uparrow$. However, if \mne\ is already canted at $T=0$ as our calculations suggest (Fig.\ \ref{fig:rotation}), it is likely not the source of the low-$T$ anomaly.

Fig.\ \ref{fig:mn-es}(c-d) shows the spectral functions for \mnsb\ at $\theta=0$ (ground state) and in the SD state. The spin moment of \mnsb\ varies from 3.82 to 4.18 $\mu_B$ between angles closest to 0 and $\pi$. Fig.\ \ref{fig:mn-es}(c) shows, in agreement with Ref.\ \onlinecite{Alling}, that \mnsb\ preserves the half-metallic gap in the ground state, although the VBM moves closer to $E_F$ compared to ideal NiMnSb. Further, in Fig.\ \ref{fig:mn-es}(d) and \ref{fig:mn-sb-dos} (dot-dashed line and inset) we see that spin disorder introduces a large spectral density inside the gap near $\Gamma$ above the VBM, which is much larger compared to \mne\ (Fig.\ \ref{fig:mn-es}(b)). These features are due to the replacement of Mn-Sb bonds by Mn-Mn; recall that Mn-Sb hybridization pushes the VBM states downward at $\Gamma$ (Fig.\ \ref{fig:bulk}(a,c)).

\begin{figure}[htb]
\includegraphics[width=0.85\columnwidth]{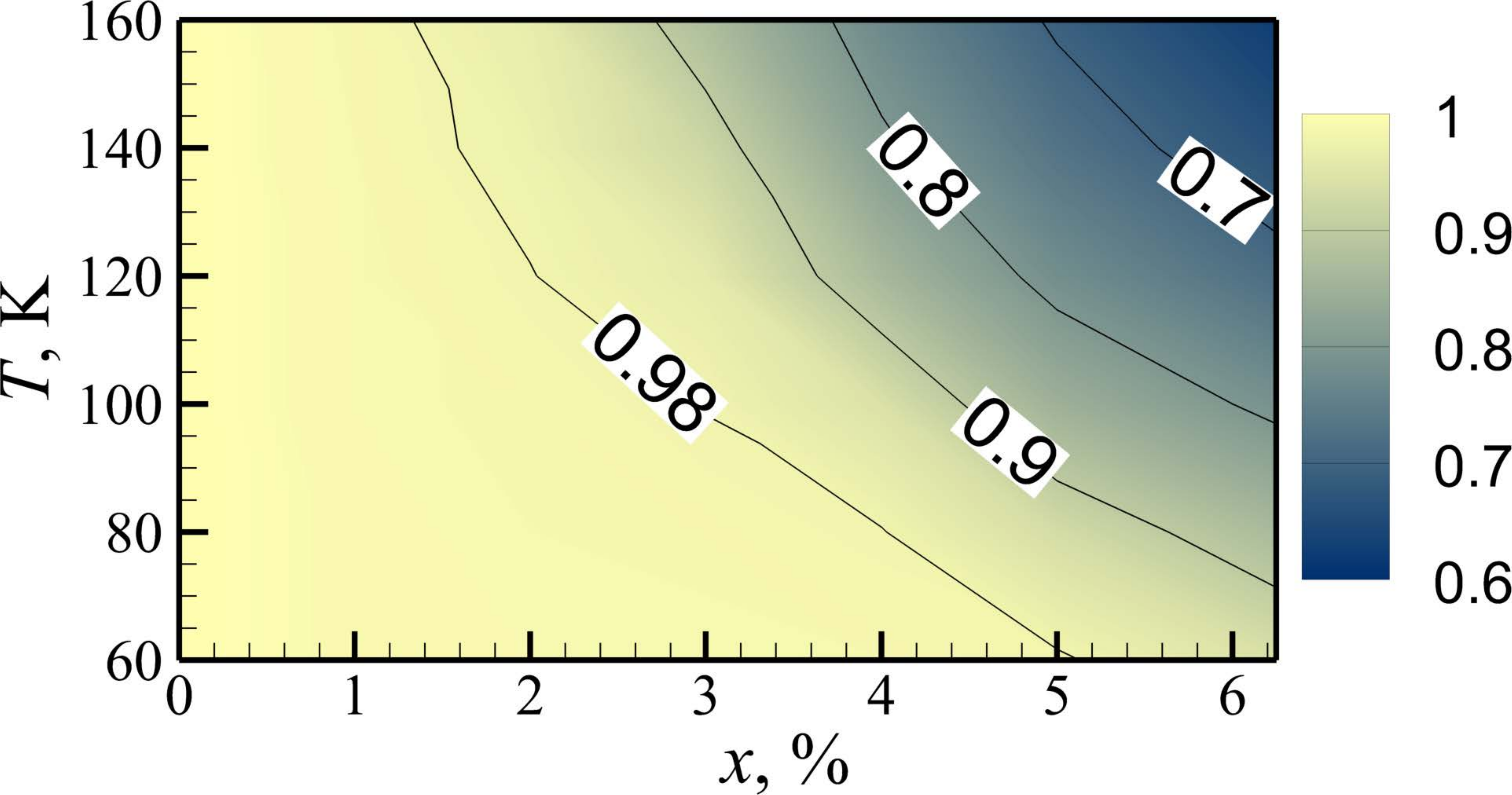}
\caption{Spectral spin polarization $P$ at $E_F$ as a function of \mnsb\ concentration $x$ and temperature.}
\label{fig:x-t-p}
\end{figure}

The inset in Fig.\ \ref{fig:mn-sb-dos} shows $N_\uparrow$ near $E_F$ for several temperatures, which were obtained using the VDLM model with the spin distribution function $p(\theta)\propto\exp[-E(\theta)/T]$ for \mnsb\ with $E(\theta)$ from VASP (Fig.\ \ref{fig:rotation}). We see that thermal spin disorder on \mnsb\ generates a considerable $N_\downarrow$ at $E_F$ at rather low temperatures; for example, $N_\downarrow\approx0.26 N_\uparrow$ at 80 K. The spectral spin polarization at $E_F$, $P=(N_\uparrow-N_\downarrow)/(N_\uparrow+N_\downarrow)$, is mapped in Fig.\ \ref{fig:x-t-p} as a function of temperature and \mnsb\ concentration. It confirms that the quick drop of $P$ occurs in the 70-100 K range near 5\% \mnsb, consistent with observations of Ref.\ \onlinecite{Wang}. At lower concentrations the crossover is shifted to higher temperatures; this is a testable feature of the thermal depolarization mechanism associated with \mnsb. 

In conclusion, we have studied the influence of thermal spin disorder and excess Mn on the electronic structure of NiMnSb. Thermal spin disorder broadens and shifts the minority-spin VBM at $\Gamma$ upwards due to the unmixing of Mn $t_{2g}$ states from Sb states. While \mnni\ is spectroscopically invisible, \mne\ and \mnsb\ have strong spectral signatures. \mnsb\ is identified as a possible source of the low-$T$ anomaly: its spin couples weakly to the bulk, and easily excited spin disorder strongly contaminates the half-metallic gap near $\Gamma$. These predictions can be tested through a combination of transport and spectroscopic measurements.

We thank Peter Dowben and Andre Petukhov for useful discussions. This work was supported by National Science Foundation through Grants DMR-1005642, DMR-1308751, and the Nebraska MRSEC (DMR-0820521). Calculations were performed utilizing the Holland Computing Center of the University of Nebraska.


\begin{thebibliography}{99}
\bibitem{basis} The basis set includes $s$, $p$, $d$ states on Ni, Mn, Sb and on an empty sphere placed at the empty site (E), as well as the $f$ states for Sb. The sphere radii are 2.62, 2.89, 2.97, and 2.48 a.\ u.\ for Ni, Mn, Sb, and E, respectively.
\bibitem{note} The structure was optimized at $\theta=0$ (\mnni\ and \mnsb) or $\theta=\pi$ (\mne), which give the lowest energy among the two collinear orientations.
\end{thebibliography}
\end{document}